\newcounter{myctr}
\def\myitem{\refstepcounter{myctr}\bibfont\noindent\ifnum\themyctr>9\else\phantom{0}\fi\hangindent17pt\themyctr.\enskip}

\documentclass{ws-ijqi}

\begin{document}

\markboth{G. Augello, D. Valenti, B. Spagnolo}
{Effects of colored noise in short overdamped Josephson junction}

\catchline{}{}{}{}{}

\title{Effects of colored noise in short overdamped Josephson junction\\
}


\author{G. Augello, D. Valenti, B. Spagnolo}


\address{Dipartimento di Fisica e Tecnologie Relative and CNISM-INFM, Unit\`{a} di Palermo\\
Group of Interdisciplinary Physics, Universit\`{a} di Palermo\\
Viale delle Scienze, I-90128 Palermo, Italy\\
augello@gip.dft.unipa.it, http://gip.dft.unipa.it}





\maketitle

\begin{history}
\received{Day Month Year}
\revised{Day Month Year}
\end{history}

\begin{abstract}

We investigate the transient dynamics of a short overdamped
Josephson junction with a periodic driving signal in the presence of
colored noise. We analyze noise induced phenomena, specifically
resonant activation and noise enhanced stability. We find that the
positions both of the minimum of RA and maximum of NES depend on the
value of the noise correlation time $\tau$$_c$. Moreover, in the
range where RA is observed, we find a non-monotonic behavior of the
mean switching time as a function of $\tau$$_c$.

\end{abstract}


\section{Introduction}    

The study of transient dynamics of Josephson junctions (JJs) in the
presence of thermal noise is of great importance for the development
of rapid single flux quantum (RSFQ)\cite{Lik} logic. RSFQ
technology is mainly based on integrated circuits composed
by superconducting quantum interference devices (SQUIDs). The high velocity of processing of
RSFQ devices is a very interesting characteristic for their use as
logic components of integrated circuits. Moreover JJs are also good candidates
as basis components of a solid state quantum computer. In addition, SQUID can be used as a storage element of the
single magnetic flux quanta (SFQs), that represent the data bits for
logic computation.\cite{Car} The commuting time of the JJs is related to the mean life time of the metastable states and to the de-coherence phenomena in quantum computation.\cite{Ber}
In this framework a great
attention was payed to the study of the mean life time of metastable states of JJs.\cite{Mal,Pank,Spa}

Noise induced effects were experimentally observed in underdamped
JJs.\cite{Yu,Sun} Resonant activation (RA)
and noise enhanced stability (NES) phenomena were found in a model of a JJ\cite{Pank,Spa}, by considering
different values of driving frequency and noise intensity. The
curves representing RA and NES showed significant modifications when
polarization current changes.

The transient dynamics of long overdamped JJs, under the influence
of white noise, was investigated through numerical simulation using
the sine-Gordon model.\cite{Fed} Due to the presence of metastable
states, noise delayed decay effects were observed in the range of
parameters useful for practical application, depending on the length
of the junctions. The results were obtained considering small noise
intensity and short junction lengths. In this range of parameters
the mean switching time (MST), that is the mean escape time from the
metastable state, shows a maximum as a function of the noise
intensity. For dimensionless lengths {\it L $>$ 5} (where {\it L =
l/$\lambda$$_J$}, with {\it l} the junction length and {\it
$\lambda$$_J$} the Josephson penetration length) the noise delayed
decay effect disappears. These results can be useful for the design
of devices based on Josephson junctions. Short overdamped JJs were studied under the influence of white noise
and fluctuating potential by numerical simulation.\cite{Gord} RA and
NES phenomena were observed. In particular it was investigated the
NES effect, finding a range of frequency in which the maximum
of MST is very prominent. Moreover the presence of NES
effect results to be influenced by the variation of the bias current.

In the present work we report the study of the transient dynamics in
short JJs. We perform numerical simulation using the Resistively
Shunted Junction model, considering short overdamped JJs under the
influence of a fluctuating potential. Noise
induced effects, such as RA and NES phenomena, in a JJ in the presence of a more realistic noise source possessing a finite correlation time (colored noise) are investigated (see Ref.~\refcite{Mar} and references therein).

\section{Model}

The Langevin equation describing the dynamics of a short overdamped
JJ under the influence of colored noise is\cite{Bar}
\begin{equation}
\frac{d\phi}{dt}=-\omega_c\frac{dU(\phi)}{d\phi}-\omega_c\zeta(t).\label{model}
\end{equation}
In Eq.~(\ref{model}) $\phi$ represents the order parameter, that is
the phase difference of the wave functions in the ground state
between left and right superconductive sides of the junction. The
characteristic frequency of the Josephson junction is
\emph{$\omega_c$=2e$R_N$$I_c$/$\hbar$}, where \emph{e} is the
electron charge, \emph{$R_N^{-1}$} is the normal conductivity,
\emph{$I_c$} is the critical current and \emph{$\hbar$=h/2$\pi$}
with \emph{h} the Plank constant. \emph{$\zeta$(t)} is an
Ornstein-Uhlenbeck (OU) process\cite{Gar}, characterized by a
correlation time $\tau_c$, representing a colored noise source.
The potential profile \emph{U($\phi$)}, is given by
\begin{equation}
U(\phi)=1-cos\phi-i(t)\phi,
\end{equation}
where \emph{$i(t)=i_0+f(t)$}, \emph{$i_0=i_b/I_c$} is the constant dimensionless bias current and \emph{f(t)=Asin$\omega$t} is the driving current with dimensionless amplitude \emph{$A=i_s/I_c$} and frequency $\omega$ (\emph{$i_b$} and \emph{$i_s$} represent the bias current and the driving current amplitude respectively).\\
The OU process of Eq.~(\ref{model}) is represented by the stochastic
differential equation\cite{Gar}
\begin{equation}
d\zeta(t)=-\frac{1}{\tau_c}\zeta(t)dt+\frac{\sqrt{\gamma}}{\tau_c}dW(t)
\end{equation}
where \emph{W(t)} is the Wiener process and \emph{$\gamma$} is the
noise intensity. The correlation function of the OU process is
\begin{equation}
\langle\zeta(t)\zeta'(t)\rangle=\frac{\gamma}{2\tau_c}e^{-\frac{{|t-t'|}}{\tau_c}}.\label{corrfun}
\end{equation}
Eq.~(\ref{model}) is a stochastic differential equation, with a
time-dependent nonlinear periodic potential, that we analyzed by
numerical simulations.
We studied the trajectories $\phi(t)$ of the particle  along the
fluctuating potential. In particular we are interested in noise
induced effects influencing the mean escape time from the metastable
state. To do this, we considered, as initial condition, the
superconductive state corresponding to one of the minima in the
potential profile, that is \emph{$\phi_0$=arcsin($i_0$)}. Then, we
calculated the MST, that is the time spent by the particle to reach
the position corresponding to the next maximum of the potential
profile. After several simulations, we obtained the behaviour of MST for
different frequency values of the fluctuating potential. The parameter of simulation are: number of realizations $N_r=10^4$, time step for integration $\delta t = 10^{-3}$, maximum value for escape time $T_{max}=10^3$. The curves
obtained show the effects of the correlation on the overall
behaviour of MST, evidencing RA and NES phenomena.

\section{Results}

We calculated the trajectory $\phi(t)$ for different values of the
characteristic parameters of the junction, such as the bias current
\emph{$i_0$} and the noise intensity $\gamma$,
keeping $\omega_c$=1. We obtained the curves of the MST and the
corresponding standard deviation (SD) vs $\omega$, and the MST vs
$\gamma$, for different values of $\tau_c$. In Fig. 1 we report the
curves of MST vs $\omega$ for white noise and colored noise, with
different values of the correlation time and noise intensity. We
found the presence of the RA phenomenon, in particular we note that
the position of the minimum in the curves depends on $\tau_c$. The
minimum is shifted towards greater values of $\omega$ when $\tau_c$
increases and this effect is more evident in correspondence of
greater values of the noise intensity (Fig. 1, right).
\begin{figure}[h]
\centerline{\psfig{file=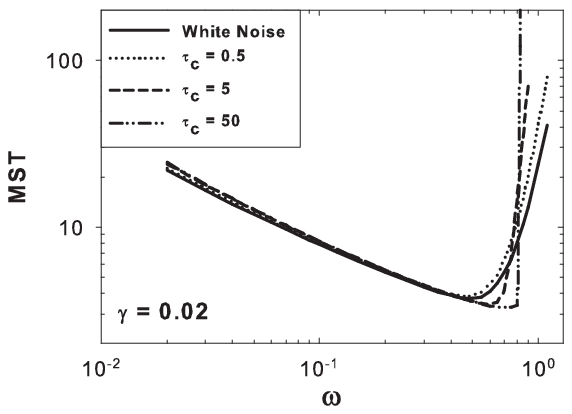}\psfig{file=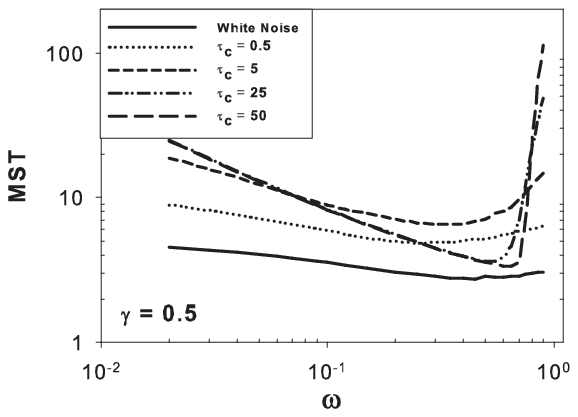}}
\vspace*{8pt} \caption{\emph{Left}: MST vs {$\omega$} for white noise and different
{$\tau_c$}, {$\gamma$}=0.02, $i_0$=0.8 and A=0.7. \emph{Right}: MST
vs {$\omega$} for white noise and different {$\tau_c$}, {$\gamma$}=0.5, $i_0$=0.8
and A=0.7.}
\end{figure}
We find a wide range of frequencies (0.3 $<$ $\omega$ $<$ 0.8) in
which a non-monotonic behaviour of MST as a function of the
correlation time is present (Fig. 1). In Fig. 2 we report the curves
representing MST vs $\tau_c$ for $\omega$=0.7 and noise
intensities $\gamma$=0.02 and $\gamma$=0.5. The values of MST, in
the presence of the white noise, diminishes when the noise intensity
grows up. When a colored noise is applied to the system, the curves
of Fig. 2 present a maximum depending on the correlation time
$\tau_c$. The maximum of MST increases for greater noise
intensities. The effect of the colored noise intensity is to enhance
MST. Therefore, by changing the value of $\tau_c$ is possible to
control the mean switching time of a JJ device. In particular, for a given
noise intensity, we find a correlation time corresponding to a
maximum of MST.
\begin{figure}[h]
\centerline{\psfig{file=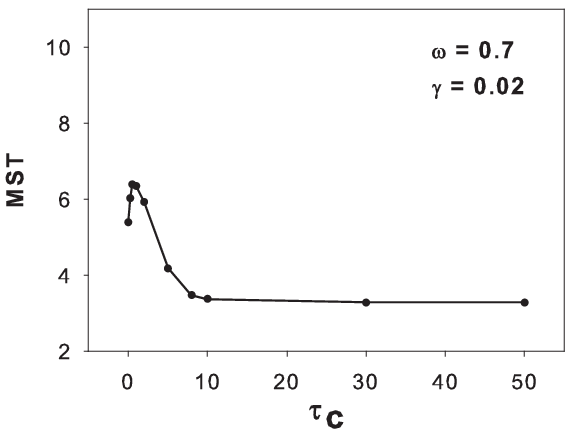}\psfig{file=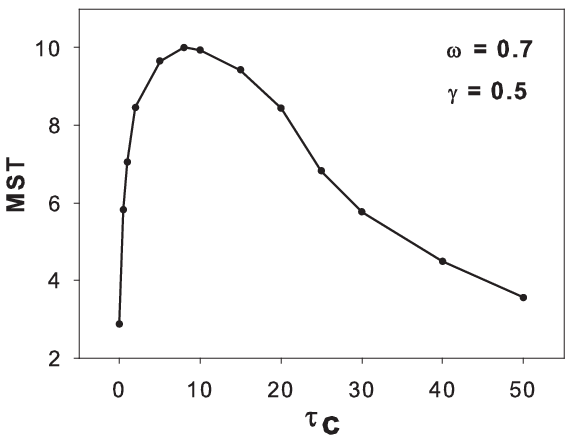}}
\vspace*{8pt} \caption{\emph{Left}: MST vs {$\tau_c$} for
{$\omega$=0.7}, {$\gamma$}=0.02, $i_0$=0.8 and A=0.7. \emph{Right}:
MST vs {$\tau_c$} for {$\omega$=0.7}, {$\gamma$}=0.5, $i_0$=0.8 and
A=0.7.}
\end{figure}
In Fig. 3 we report the curves for MST and SD as a function of
$\omega$. We note a range of frequency (0.2 $<$ $\omega$ $<$ 0.8) in
which MST and SD present a minimum.\cite{Pank}
\begin{figure}[h]
\centerline{\psfig{file=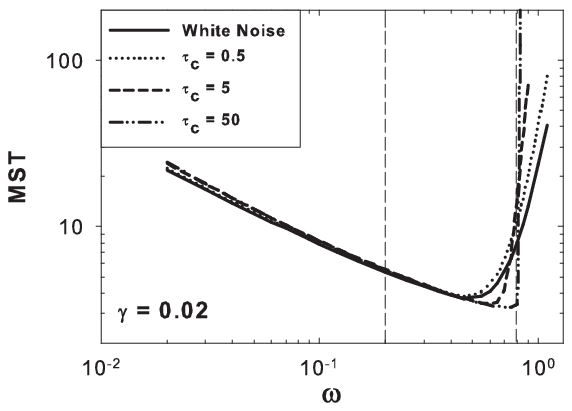}\psfig{file=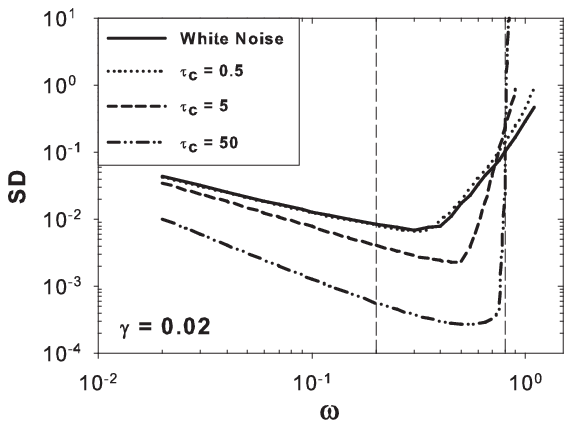}}\label{f2}
\vspace*{8pt} \caption{\emph{Left}: MST vs {$\omega$} for different
{$\tau_c$}, {$\gamma$}=0.02, $i_0$=0.8 and A=0.7. \emph{Right}:
Standard deviation (SD) vs {$\omega$}.}
\end{figure}

We find also that the RA phenomenon, in the presence of colored
noise, presents a scaling effect depending on the values of the
correlation time. In Eq.~(\ref{corrfun}) for correlation time
greater than the characteristic time scale of the system,
$\tau_c$$\gg$$|$t-t$'$$|$, we obtain
$\langle$$\zeta$(t)$\zeta$(t$'$)$\rangle$$\approx${$\gamma_{colored}$}/{2$\tau_c$}.
By comparison with the correlation function of the white noise,
$\langle$$\xi$(t)$\xi$(t$'$)$\rangle$={$\gamma_{white}$}$\delta(t-t')$,
we note that the effective intensity of the colored noise is
equivalent, in this approximation, to the intensity of the white noise, scaled by a factor
1/2$\tau_c$:
\begin{equation}
\gamma_{white}=\frac{\gamma_{colored}}{2\tau_c}
\end{equation}
To verify this result we considered the effect on the system of a
white noise with intensity $\gamma_{white}$=0.005 and a colored
noise with intensity $\gamma_{colored}$=0.5, setting $\tau_c$=50. In
Fig. 4, on the left, we report the curves of MST vs $\omega$
calculated for these two values of noise intensity. We see that for
MST less than 50 there is a good agreement between the curves. In
this conditions the time evolution is faster than the correlation
time and the system is not affected by the correlation.
\begin{figure}[h]
\centerline{\psfig{file=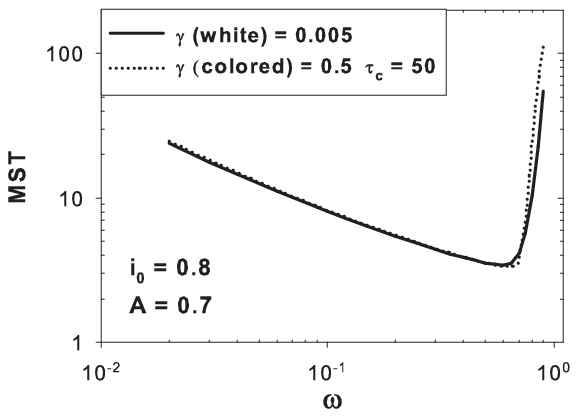}\psfig{file=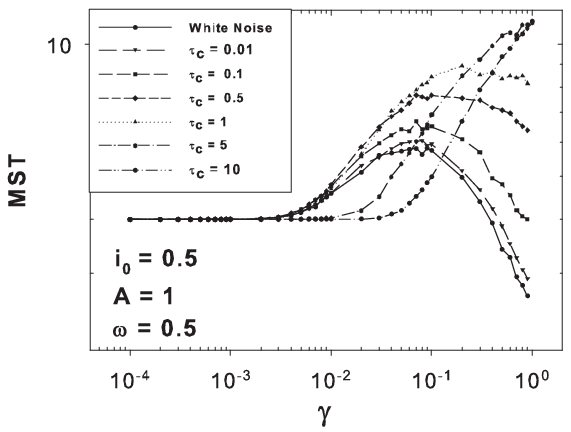}}
\vspace*{8pt} \caption{\emph{Left}: MST vs $\omega$ with $i_0$=0.8,
A=0.7: white noise (line) and colored noise with $\tau_c$=50 (dots).
\emph{Right}: MST vs {$\gamma$} for white noise and colored noise
with different value of $\tau_c$, $\omega$=0.5, A=1, $i_0$=0.5.}
\end{figure}

We investigated the NES phenomenon by studying the curves of MST as
a function of the noise intensity, for different values of the
correlation time (Fig. 4, right). The NES phenomenon had been
already found in short JJs in the presence of white
noise\cite{Gord}.
Here we found a range of values of the correlation time,
0.01$<$$\tau_c$$<$1, in which the curves present a non-monotonic
behaviour. For $\tau_c$=5,10, the non-monotonic behaviour
disappears. As already noted in Fig. 1, we observe the presence of a
non-monotonic behaviour of MST as a function of $\tau_c$.
One of the main results of this work concerns the optimum range of the system parameter values in which the MST and its SD are minimized. Specifically, for 0.3$<$$\omega$$<$0.7, $\gamma\approx0.02$ and $\tau_c>5$, the MST and its SD take the lower values and the enhancement of MST, due to the NES effect, is vanishing (see Figs. 3 and 4-right). As a consequence  the efficiency of JJ device, as a logic component, in this parameter range increases.

\section{Conclusions}

We studied the transient dynamics of Josephson junctions under the
influence of colored noise and driving potential oscillating with
frequency $\omega$. We found the presence of noise induced effects
such as resonant activation and noise enhanced stability phenomena.
RA is influenced by the presence of colored
noise: varying the correlation time $\tau_c$ affects
the position of the minimum of MST. In particular, we noted a
non-monotonic behaviour of MST as a function of $\tau_c$. We found a
range in which both MST and the corresponding SD present a minimum,
as a function of $\omega$, for different values of the correlation
time. Moreover we found a scaling effect of the colored noise when
large values of the correlation time are considered. Finally we
studied the curves of the mean switching time as a function of the
noise intensity, finding that the correlation time affects the noise
enhanced stability phenomenon.

\section*{Acknowledgments}

The authors acknowledge the financial support of MUR. This work
makes use of results produced by the PI2S2 Project managed by the
Consorzio COMETA, a project co-funded by the Italian Ministry of
University and Research (MUR) within the Piano Operativo Nazionale
"Ricerca Scientifica, Sviluppo Tecnologico, Alta Formazione" (PON
2000-2006). More information is available at http://www.pi2s2.it and
http://www.consorzio-cometa.it.

\end{document}